\documentclass[12pt]{iopart}

\usepackage{graphicx}
\begin{document}

\title[Collective Flow Measurements from the PHENIX Experiment]
{Collective Flow Measurements from the PHENIX Experiment}

\author{ShinIchi Esumi (for the PHENIX Collaboration\footnote{A list of 
members of the PHENIX Collaboration can be found at the end of this issue.}) }
\address{Inst. of Physics, Univ. of Tsukuba,\\
Tennno-dai 1-1-1, Tsukuba, Ibaraki 305-8571, Japan}
\ead{esumi@sakura.cc.tsukuba.ac.jp}
\begin{abstract}
Recent collective flow measurements including higher moment
event anisotropy from the PHENIX experiment are presented,
and the particle type, beam energy dependence and the relation 
with jet modification are discussed. 
The measured higher order event anisotropy with event plane 
defined at forward rapidities and the long range correlation 
with large $\eta$ gaps are both consistent with initial 
geometrical fluctuation of the participating nuclei. 
In 200 GeV Au+Au collisions, higher order event anisotropy, especially 
simultaneous description of v$_2$ and v$_3$, is found to give 
an additional constraining power on initial geometrical condition 
and viscosity in the hydrodynamic calculations. v$_2$, v$_3$ and 
v$_4$ are almost unchanged down to the lower colliding energy 
at 39 GeV in Au+Au. 
The measured two particle correlation with subtraction of the 
measured v$_n$ parameters shows a significant effect on the 
shape and yield in the associate particle $\Delta\phi$ distribution 
with respect to the azimuthal direction of trigger particles. 
However some medium responses from jet suppression or jet 
modification seems to be observed. 
Direct photon v$_2$ has been measured in 200 GeV Au+Au 
collisions. The measured v$_2$ is found to be small at high 
p$_{\rm T}$ as expected from non-suppressed direct photon 
R$_{\rm AA}$ $\simeq$ 1, which can be understood as being 
dominated by prompt photons from initial hard scattering. 
On the other hand, at lower p$_{\rm T}$ $<$ 4 GeV/c it is 
found to be significantly larger than zero, which is 
comparable to other hadron v$_2$, where thermal photons 
are observed. 
\end{abstract}


\section{Higher order event anisotropy measurements}

\begin{figure}[h]
\begin{center}
\includegraphics[width=12cm]{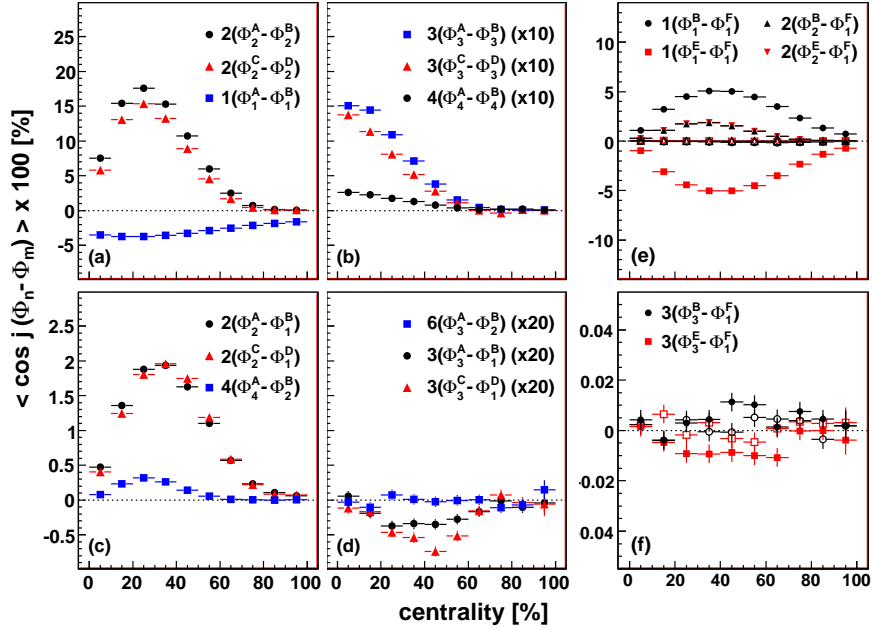}
\caption{higher order event plane correlation between 
same and different harmonics and between different rapidities, 
where the subscript n=1$\sim$4 and superscript X=A$\sim$F indicated on 
$\Phi_{\rm n}^{\rm X}$ in the figure are the order of harmonics and 
the index of the following sub-detectors, respectively; 
A:RXN$^{(+)}$, B:BBC$^{(-)}$, C:MPC$^{(+)}$, D:MPC$^{(-)}$, 
E:BBC$^{(+)}$ and F:ZDC$^{(+/-)}$, where $(+)$ or $(-)$ 
indicates the positive or the negative rapidity. The descriptions 
of the sub-detectors and their $\eta$ acceptances are given in 
the text.
\label{Esumi_fig0}}
\end{center}
\end{figure}

In order to measure the higher order event anisotropy, the 
event planes from the higher moments at various rapidities 
are defined with various reaction plane detectors; 
the Reaction Plane detector (RXN: $|\eta|$ = 1.0 - 2.8), 
the Muon Piston Calorimeter (MPC: $|\eta|$ = 3.1 - 3.7), 
the Beam-Beam Counter (BBC: $|\eta|$ = 3.1 - 3.9) and 
the Zero Degree Calorimeter (ZDC: $|\eta|$ $>$ 6.5). 
Fig.\ref{Esumi_fig0} shows the event plane correlations 
between forward and backward rapidities for the same moments 
in panels (a) and (b), for the different moments in panels (c) 
and (d). The significant positive correlations between the same 
order higher moments are visible in panel (b), which is consistent 
with the initial geometrical fluctuation\cite{ref0}. 
A weak negative correlation between 1$^{\rm st}$ and 3$^{\rm rd}$ 
moment event planes can be an indication of 
rapidity anti-symmetric v$_3$ contribution as seen in panel (d), 
which is further confirmed by the correlation measurements 
with respect to the 1$^{\rm st}$ moment event plane from 
the spectators shown in panels (e) and (f), 
where the weak anti-symmetric v$_3$ 
contribution shows the same sign as v$_1$ at the same rapidity. 
No visible correlation between 2$^{\rm nd}$ and 3$^{\rm rd}$ harmonics 
seen in panel (d), this also indicates that the presence of geometrical 
fluctuation is the dominant effect.

\begin{figure}[h]
\begin{center}
\includegraphics[width=12cm]{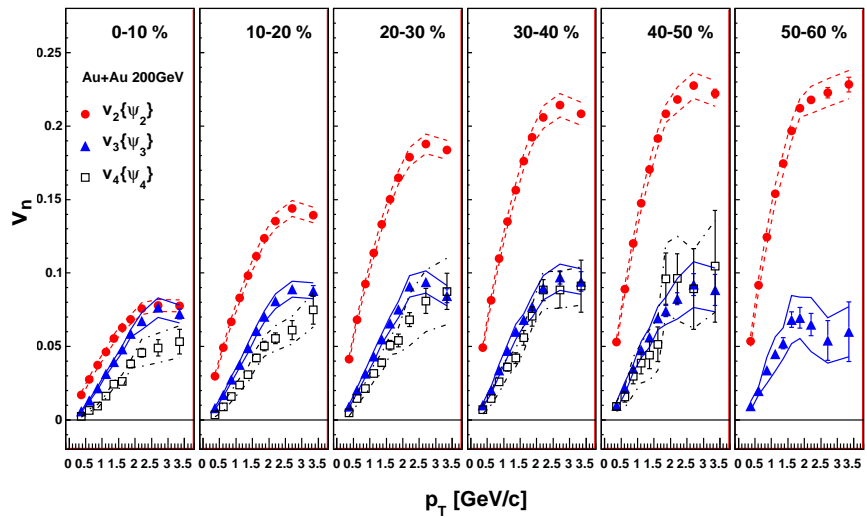}
\caption{measured v$_n$ parameters as a function of p$_{\rm T}$ 
for different centrality slices at 200 GeV Au+Au collisions
\label{Esumi_fig1}}
\end{center}
\end{figure}

\begin{figure}[h]
\begin{center}
\includegraphics[width=10cm]{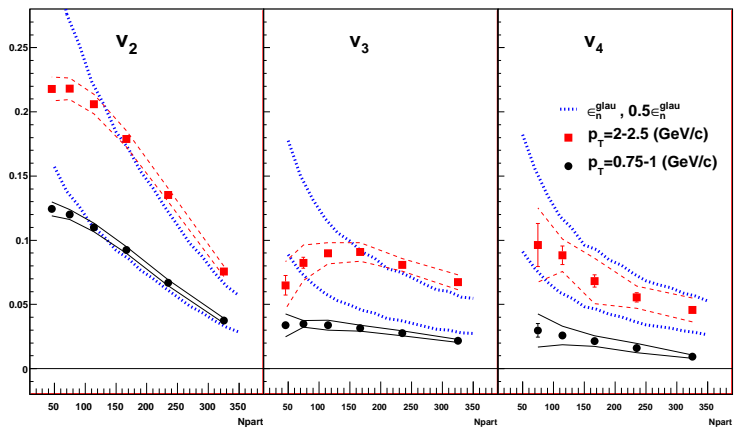}
\caption{measured v$_n$ parameters as a function of N$_{\rm part}$
(for two different p$_{\rm T}$ bins) at 200 GeV Au+Au collisions 
\label{Esumi_fig2}}
\end{center}
\end{figure}

The measured v$_2$, v$_3$ and v$_4$ with respect to the same order 
event plane $\Psi_n$ defined at forward rapidity in RXN are shown 
as a function of p$_{\rm T}$ for several centrality slices in each 
panel in Fig.\ref{Esumi_fig1} \cite{ref1}
and as a function of N$_{\rm part}$ for two p$_{\rm T}$ bins in 
Fig.\ref{Esumi_fig2} \cite{ref1}, N$_{\rm part}$ dependence is compared with 
a generalized eccentricity up to the higher moments. Significant 
higher moments compared to v$_2$ are observed for central collisions, 
and smaller centrality dependence is seen especially in v$_3$. One 
should also note that this v$_4$($\Psi_4$) seems to be 
larger than the previously measured v$_4$($\Psi_2$) \cite{ref2}, 
because of the additional fluctuation contribution in $\Psi_4$ angle 
with respect to $\Psi_2$ angle.

\begin{figure}[h]
\begin{center}
\includegraphics[width=11cm]{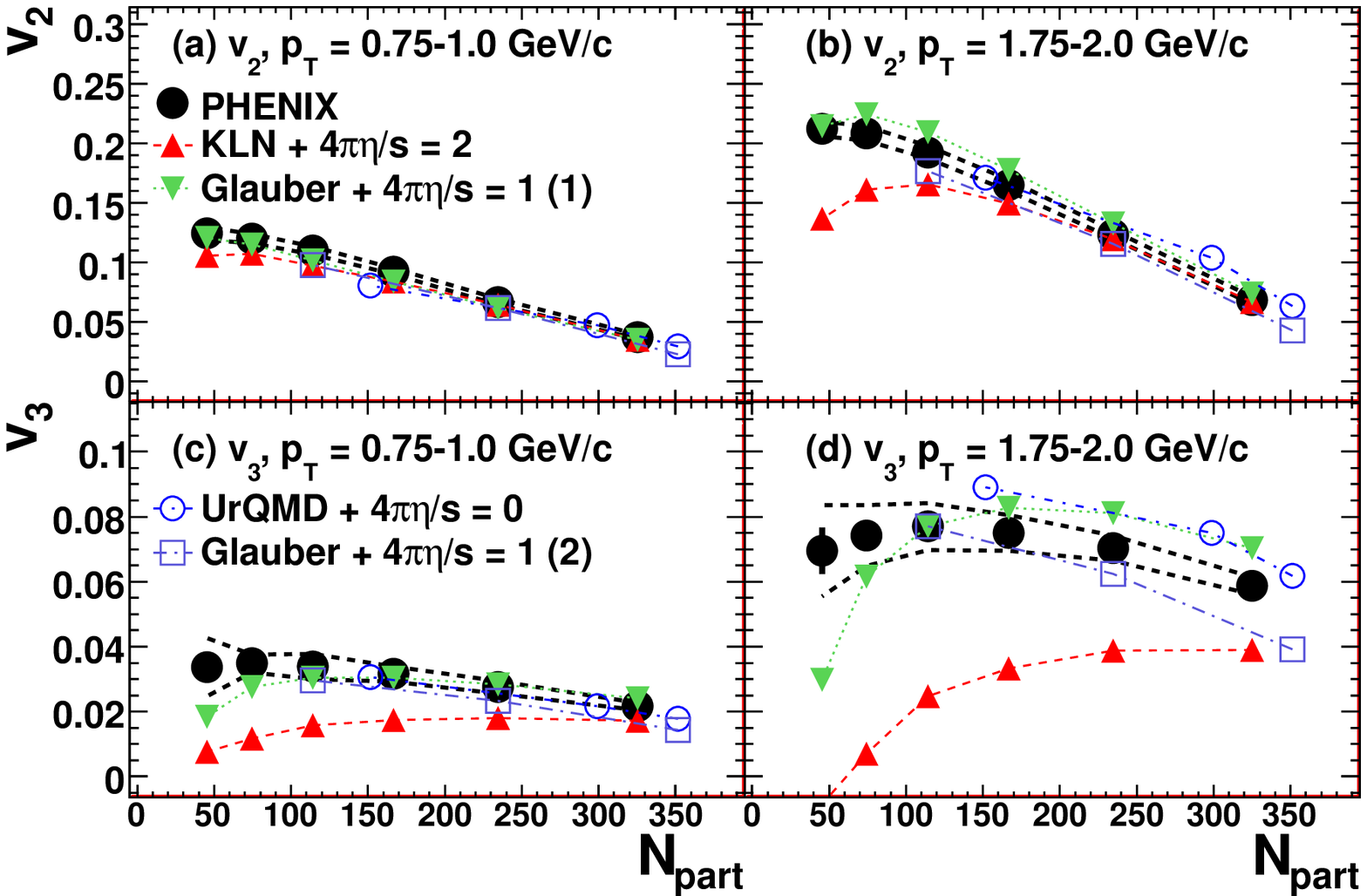}
\caption{comparison of v$_2$ and v$_3$ with various model calculations 
\label{Esumi_fig3}}
\end{center}
\end{figure}

\begin{figure}[h]
\begin{center}
\includegraphics[width=8cm]{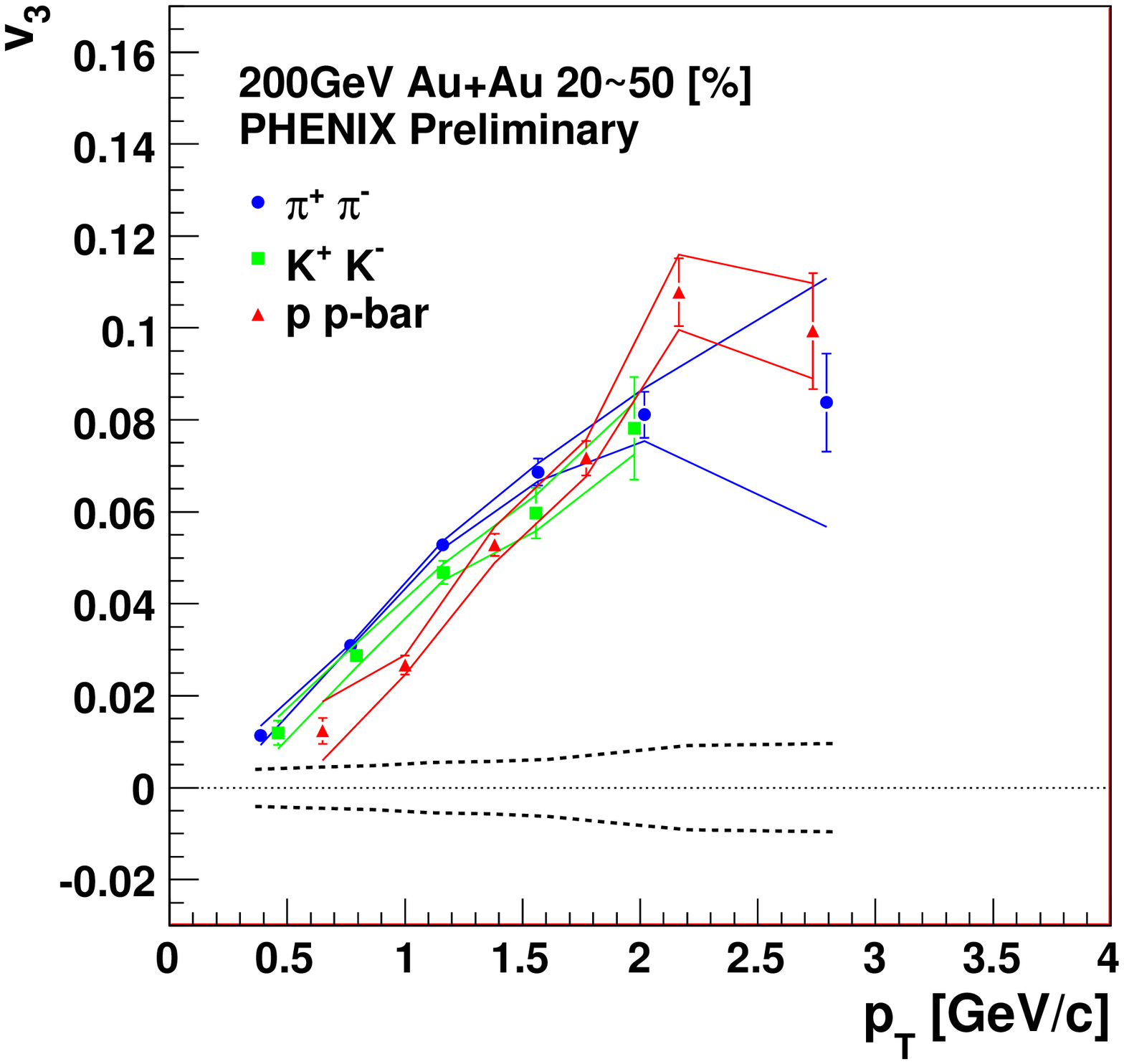}
\caption{Identified hadron v$_3$ as a function of p$_{\rm T}$
at 200 GeV Au+Au collisions \label{Esumi_fig4}}
\end{center}
\end{figure}

The measured v$_2$ and v$_3$ are compared with various model 
calculations in Fig.\ref{Esumi_fig3}. 
Since the initial generalized eccentricities 
$\epsilon_n$ between Glauber model and CGC-KLN model have 
opposite trends for 2$^{\rm nd}$ and 3$^{\rm rd}$ moments because 
of different fluctuation contribution with respect to the initial 
overlapping geometry, 
the shear viscosity over entropy ratio $\eta/s$ used 
in hydrodynamic calculations will be more constrained by 
simultaneous fitting to the measured v$_2$ and v$_3$, compared 
with the previous v$_2$ alone fitting. The existing hydrodynamic 
calculations seem to describe 
the experimental measurements using the Glauber type initial 
condition with a smaller shear viscosity and the other calculations 
with the MC-KLN type initial condition with a larger shear viscosity 
are disfavored by the data\cite{ref3,ref4}.  

Identified hadron v$_3$ is shown as a function of p$_{\rm T}$ 
for $\pi^{+-}$, $K^{+-}$ and $p$+$\overline{p}$ in Fig.\ref{Esumi_fig4}. 
The similar mass ordering of v$_3$ compared with v$_2$ is 
observed at relatively lower p$_{\rm T}$ region as one might expect
from the radial flow picture from hydrodynamic expansion, while 
at intermediate p$_{\rm T}$ region, the similar baryon - meson 
splitting in v$_3$ given by the number of constituent quarks could 
possibly be seen just like the quark number scaling of v$_2$, which 
might be related to the partonic degrees of freedom during the 
v$_3$ evolution. Both observations at least tell us that the 
v$_3$, which is originated from the initial geometrical fluctuation, 
seems to reflect the hydrodynamic evolution as triangular expansion. 

The higher harmonics event anisotropy v$_n$ measurements for charged 
particles and the identified hadron v$_2$ measurements has also been 
done for both 62 GeV and 39 GeV Au+Au collisions, the observed v$_n$ and 
v$_2$ results for given p$_{\rm T}$ and centrality show very similar 
results to those of 200 GeV Au+Au collisions, which would indicate 
that the hydrodynamical properties are mostly unchanged within 
the collision energy region 39-200 GeV\cite{ref4a}. 

\section{Implication on two-particle correlation measurements}

\begin{figure}[h]
\begin{center}
\includegraphics[width=12cm]{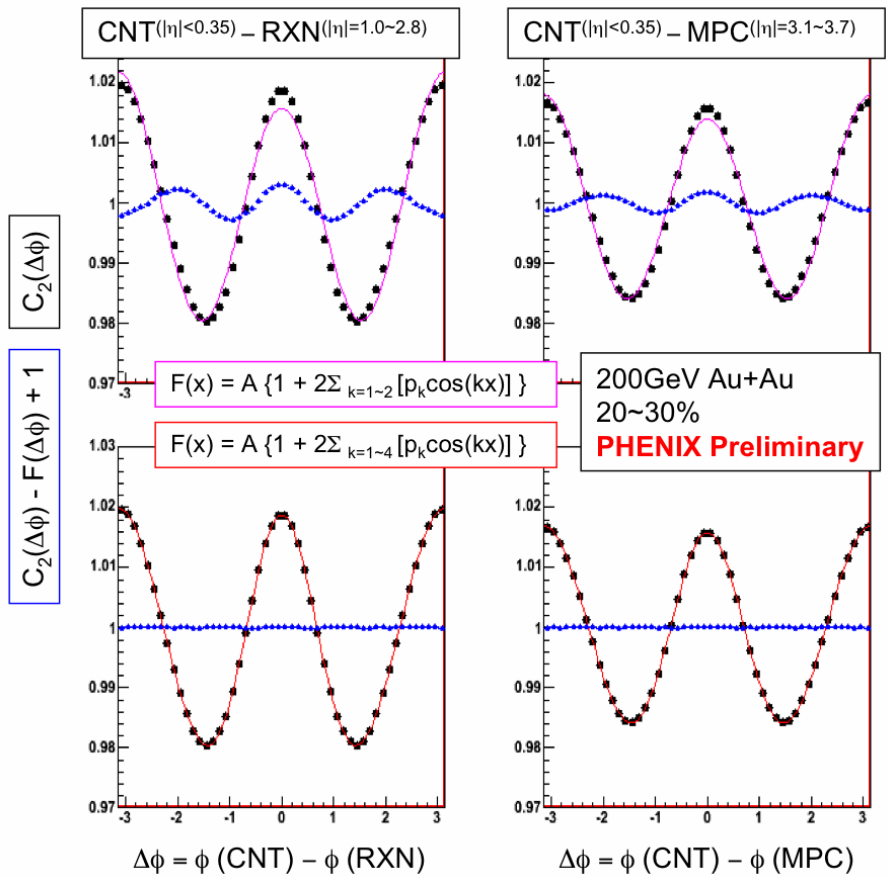}
\caption{two particle azimuthal correlation between forward and 
central rapidities \label{Esumi_fig5}}
\end{center}
\end{figure}

Fig.\ref{Esumi_fig5} shows the two particle azimuthal correlation 
between forward and central rapidities, where the left panels are 
with RXN and the right panels are with MPC for forward rapidity. 
The both top and bottom panels show the same data points (black) 
and the data are fitted with the Fourier function shown as red curves 
up to the 2$^{\rm nd}$ order for top panels and up to the 4$^{\rm th}$ 
order for the bottom panels. 
The difference between data points and fitted curve is shown as 
blue points, where the correlation is well described by including 
up to the 4$^{\rm th}$ order as shown in the bottom panels, 
while the differences shown in top panels show 
significant 3$^{\rm rd}$ moment signal, which has previously been considered 
as the near-side ridge and away-side double-peak like structure. 
These fitted parameters can then be used to extract the higher order 
event anisotropy and has been confirmed to be consistent with the 
results from the event plane method. One has to keep in mind that 
the same experimental observations (near- and away-side structures) 
are just expressed in a different way by the higher order harmonic 
parameters v$_n$. 

\begin{figure}[h]
\begin{center}
\includegraphics[width=11cm]{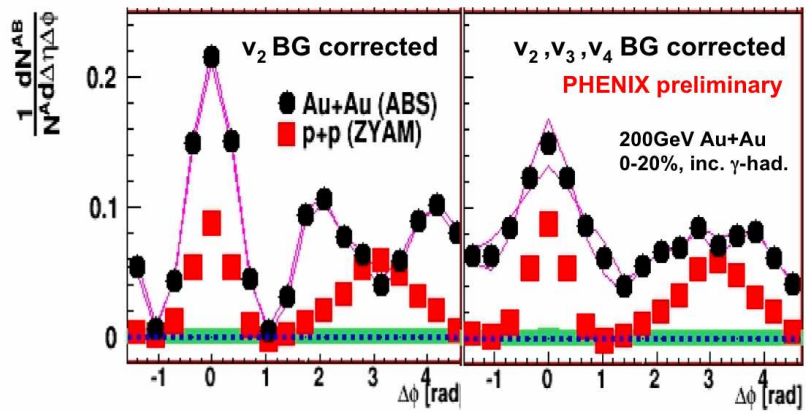}
\caption{two particle azimuthal correlation within the central 
rapidities \label{Esumi_fig6}}
\end{center}
\end{figure}

Fig.\ref{Esumi_fig6} shows the inclusive $\gamma$-hadron correlation
measured within the central rapidity region ($|\eta|$ $<$ 0.35) in 
central Au+Au collisions compared to p+p collisions\cite{ref4b}. 
The left panel shows the result with the v$_2$ BG 
correction, while the right panel shows the corrected results with 
the measured v$_2$, v$_3$ and v$_4$ parameters. The jet associated 
yields for Au+Au collisions are determined by the absolute 
normalization method without the ZYAM assumption.  
Although the away-side double-peak like structure has been 
significantly reduced with including the higher moment parameters 
v$_n$ in the BG correction, there seems to be some remaining effects 
from the jet suppression and/or modification in terms of broadening 
and enhancement in both near- and away-side shape relative to the 
shape in p+p collisions. 

\section{Direct photon elliptic event anisotropy measurements}

\begin{figure}[h]
\begin{center}
\includegraphics[width=12cm]{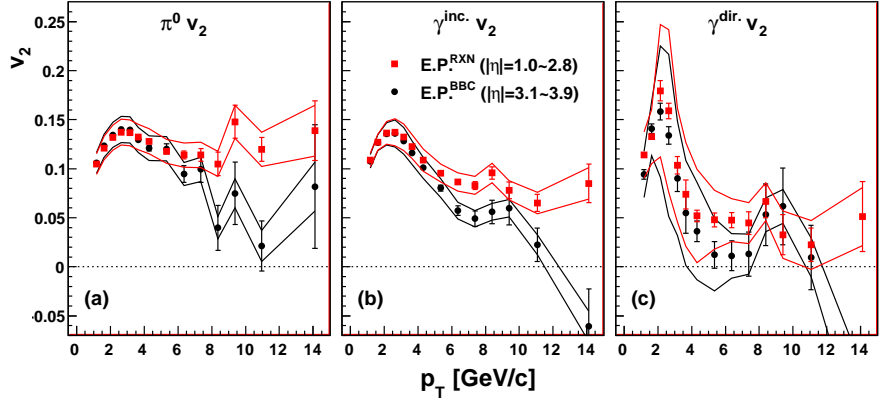}
\caption{$\pi^0$, $\gamma^{\rm inc.}$ and $\gamma^{\rm dir.}$ v$_2$ 
parameters as a function of p$_{\rm T}$ at 200 GeV Au+Au minimum bias 
collisions\label{Esumi_fig7}}
\end{center}
\end{figure}

\begin{figure}[h]
\begin{center}
\includegraphics[width=12cm]{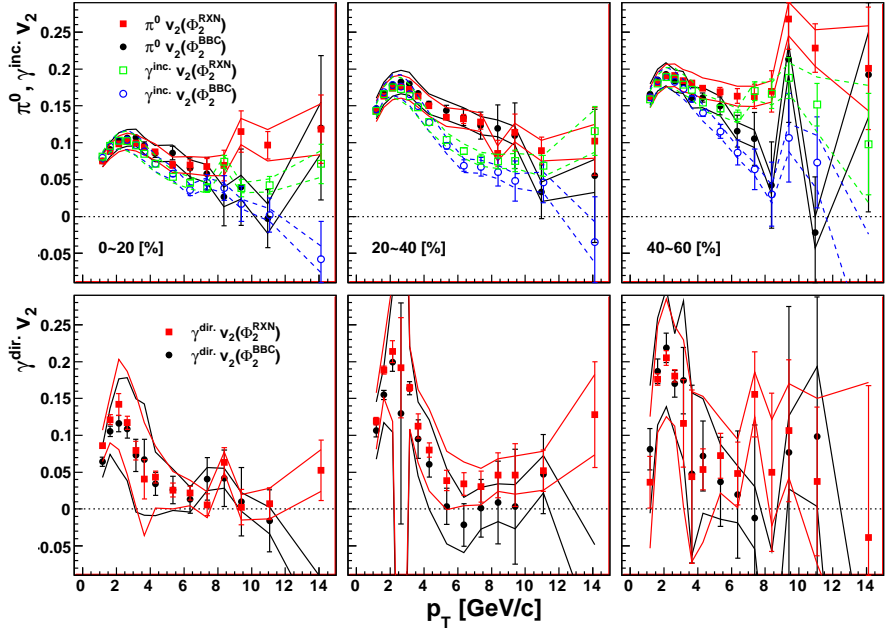}
\caption{$\pi^0$, $\gamma^{\rm inc.}$ and $\gamma^{\rm dir.}$ v$_2$ 
parameters as a function of p$_{\rm T}$ for three centrality 
selections at 200 GeV Au+Au collisions\label{Esumi_fig8}}
\end{center}
\end{figure}

The direct photon carries various different contributions; (1) prompt photon 
from initial hard scattering at high p$_{\rm T}$, which is expected to 
give zero v$_2$, (2) fragmentation photon after the parton energy loss, 
which should follow the hadron as positive v$_2$, (3) jet conversion photon 
and Bremstrahlung photon, which would have negative v$_2$, since this 
is a part of energy loss itself, (4) thermal photon, which would carry 
the history of the dynamical evolution of quark gluon plasma. 
The direct photon v$_2$ has been determined by the measured $\pi^0$ 
and inclusive photon v$_2$ and relative direct photon yield over the 
total inclusive photon yield including various hadronic decay photons. 
This measurement has been extremely difficult because of the relatively 
small direct photon signal especially at low p$_{\rm T}$ region. Using 
the virtual photon measurement\cite{ref5,ref6}, 
the accuracy of the relative direct photon 
yield determination over the total inclusive photon yield has been greatly 
improved. This improvement has given a significant reduction 
of systematic error on the extracted direct photon v$_2$ value below 
4 GeV/c in p$_{\rm T}$ in the following figures. 
The measured $\pi^0$, $\gamma^{inc.}$ and 
$\gamma^{dir.}$ v$_2$ are shown in Fig.\ref{Esumi_fig7} and 
Fig.\ref{Esumi_fig8} as a function of p$_{\rm T}$ in 200 GeV 
Au+Au collisions for minimum bias and for three centrality bins, 
respectively\cite{ref7}. 
The data are shown with two event planes defined at two different 
rapidities. The difference between two event planes is clearly seen for 
$\pi^0$ and $\gamma^{inc.}$ v$_2$ at higher p$_{\rm T}$ region expected 
from the possible non-flow contributions, however the difference 
in the $\gamma^{dir.}$ v$_2$ is smaller than the systematic error bands. 

\begin{figure}[h]
\begin{center}
\includegraphics[width=12cm]{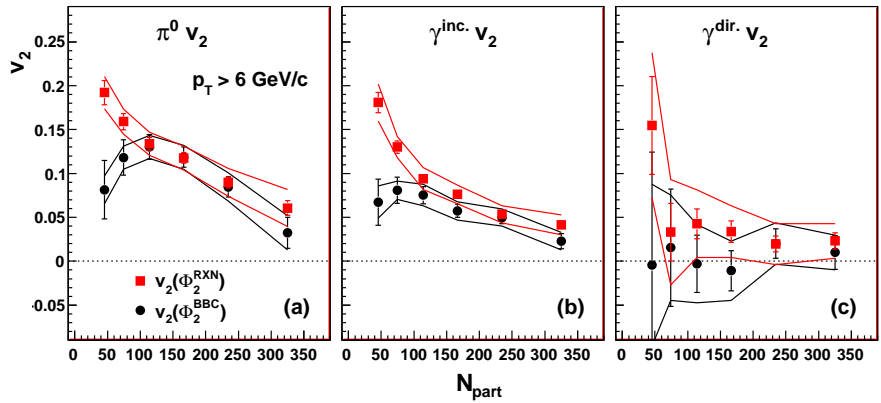}
\caption{$\pi^0$, $\gamma^{\rm inc.}$ and $\gamma^{\rm dir.}$ v$_2$ 
parameters (integrated above 6 GeV/c) as a function of N$_{\rm part}$ 
at 200 GeV Au+Au collisions \label{Esumi_fig9}}
\end{center}
\end{figure}

At the higher p$_{\rm T}$ region, the direct photon v$_2$ is consistent 
with zero as also seen in Fig.\ref{Esumi_fig9}, which confirms 
the previous direct photon R$_{\rm AA}$ measurement to be about 1
because of the dominant prompt photon production. On the other hand, 
at the lower p$_{\rm T}$ region as shown in Fig.\ref{Esumi_fig7} and 
Fig.\ref{Esumi_fig8}, the direct photon v$_2$ is found to 
be significantly larger than zero and similar to the hadron v$_2$.  
Various model calculations assuming entire history of thermal photon 
emissions which includes a significant early time (small v$_2$) 
contribution have failed to explain the experimental measurement.

\section{Conclusion}

The recent results on the collective flow measurements from the PHENIX 
experiment are presented. The higher order event anisotropy has been 
measured to be consistent with initial geometrical fluctuation followed 
by hydrodynamic expansion. The initial condition and shear viscosity in 
hydrodynamic calculations are found to be more constrained by these 
measurements. This also strongly influences on the understanding of 
the extracted shape of two particle correlation from the jet-medium 
interaction. The direct photon elliptic event anisotropy v$_2$ has 
been observed to be small as expected from the prompt photon production 
at high p$_{\rm T}$, while it has been observed to be as large as the 
hadronic v$_2$ at lower p$_{\rm T}$, where thermal photons are seen.

\end{document}